\documentclass[a4paper,12pt]{article}
\usepackage[T2A]{fontenc}
\usepackage[utf8]{inputenc}
\usepackage[english]{babel}
\usepackage[dvips]{graphicx}
\usepackage{caption}
\graphicspath{{images/}}
\usepackage[toc,title,page]{appendix}
\usepackage{verbatim}

\usepackage[hcentering, bindingoffset = 10mm, right = 15 mm, left = 10 mm, top=20 mm, bottom = 20 mm]{geometry} 

\DeclareGraphicsExtensions{.HEIC,.pdf,.png,.jpg} 
\usepackage{mathtext,cite, tikz, amsmath, amsfonts, amssymb, amsthm, mathtools, multirow, lipsum, amsmath, amstext, subcaption, upgreek, wrapfig, adjustbox, icomma, amssymb, enumerate, indentfirst, float, mathrsfs, hyperref, gensymb, chngpage, pgfplots, hhline, tabularx, xfrac, tikz-feynman, physics}

\title{On the determination of the relative probability of $\mathit{\Upsilon}(5S) \rightarrow B^{(*)}_s\bar B^{(*)}_s$ decays}

\author{\bfseries A.E. Bondar \\ 
\textit{
Budker Institute of Nuclear Physics of
SB RAS,
Novosibirsk, 630090, Russia} \\
\textit{Novosibirsk State University, 630090 Novosibirsk, Russia}\\ [2ex]
\bfseries E.K. Karkaryan   \footnote{karkaryan@bk.ru}\\
 \textit{I.E. Tamm Department of Theoretical Physics, Lebedev Physical Institute,}\\
 \textit{53 Leninskiy Prospekt, Moscow, 119991, Russia}\\[2ex] 
\bfseries A.A. Simovonian \\ 
\textit{Moscow Institute of Physics and Technology, 
Dolgoprudny 141700, Russia} \\[2ex]
\bfseries M.I. Vysotsky\\
\textit{I.E. Tamm Department of Theoretical Physics, Lebedev Physical Institute,}\\
 \textit{53 Leninskiy Prospekt, Moscow, 119991, Russia}}

\begin{document}
\maketitle
\begin{abstract}
	Semileptonic decays of the $B \overline{B}$ pairs produced in the $\mathit{\Upsilon}(5S)$ can be used to find the relative probability of $\mathit{\Upsilon}(5S) \rightarrow B_s \overline{B}_s$ decays. This could be achieved by the study of time dependence of $B$-meson decays to the leptons of equal and opposite signs. 
  \end{abstract}
\section{Introduction}

Now, when the Standard Model is crowned by the discovery of the Higgs boson, the next step is detecting the signals of «New Physics»(NP). One of the possible paths in this direction is the study of rare decays, in particular strongly suppressed decay $B_s \rightarrow \mu^+\mu^-$ \cite{Fleischer:2010ay}. 
To get the theoretical prediction for the branching ratio of this decay in the Standard Model we follow paper \cite{Bobeth:2021cxm} and extract it from the ratio $\mathcal{B}(B_s \to \mu^+\mu^-)/\Delta M_s$. In this way we get $\mathcal{B}(B_s \rightarrow \mu^+\mu^-) = (3.63^{+0.15}_{-0.10})\times 10^{-9}$. It is well known that this value can be significantly changed by NP \cite{Buras:2009if}. 

The corresponding branching ratio was measured at LHC: $\mathcal{B}(B_s \rightarrow \mu^+\mu^-) = (3.34 \pm 0.27)\times 10^{-9}$\cite{pdglive}. It is highly desirable to measure this branching ratio with approximately one order of magnitude smaller error approaching in this way theoretical accuracy and making definite prediction concerning the possible contribution of NP in this decay. Uncertainty in the number of $B_s$-mesons produced at LHC does not allow to measure $\mathcal{B}(B_s \rightarrow \mu^+\mu^-)$ with the desired small uncertainty.

In \cite{LHCb:2021qbv} the ratio of $B_s$ and $B^0$ fragmentation fractions $f_s/f_d$ at the LHC was determined with $3\%$ uncertainty (these fragmentation fractions are the probabilities for a $b$-quark to hadronise into $B_s$ or $B^0$ meson). It can be used for an accurate extraction of the $B_s \to \mu^+\mu^-$ branching ratio. However from the same measurements the following result was obtained: $\mathcal{B}(B_s \to D_s\pi^+)/\mathcal{B}(B^0 \to D^-\pi^+) = 1.18\pm0.04$. 
According to the calculations performed in \cite{Dib:2023vot}, this ratio is expected to be equal to $1.02 \pm 0.05$, which differs by more than two standard deviations from the experimental result. The measurement of the cross section of $B_s$ production at the LHC is highly complicated and model dependent procedure. Thus an independent test by a completely different method is highly desirable.

The fraction of $B_s\overline{B}_s$ events at $\mathit{\Upsilon}(5S)$ was measured in \cite{Belle:2023yfw}: $f_s = (22.0^{+2.0}_{-2.1})\%$. The main source of the error is the systematic uncertainty in the branching of the $B_s \rightarrow D^{\pm}_s X$ decay. 

The model independent method of the measurement of the value of $f_s$ was suggested in \cite{Sia:2006cq}~\footnote{ The value of $f_s$ in article \cite{Belle:2023yfw} is defined as the fraction of $B_s\overline{B}_s$ events relative to any events of $b\overline{b}$ quark production while in article \cite{Sia:2006cq} in this relation only the events with $B$ mesons pairs are used. It should be noted here that the values of $f_s$ defined in \cite{Belle:2023yfw} and \cite{Sia:2006cq} are somewhat different, since $\Upsilon(5S)$ can decay not only into $B$ mesons, but also into lighter bottomonia \cite{ParticleDataGroup:2022pth}. For further discussion, this difference does not seem significant. }. It uses the theoretical prediction, made in \cite{Bigi:2011gf} that semileptonic widths of $B^0$, $B^+$ and $B_s$ are equal within $1\%$ accuracy. One can determine the value of $f_s$ from the ratio of the number of semileptonic decays of both $B$-mesons into the leptons of equal signs to that of total number of semileptonic decays of both $B$-mesons. 
One of the main limiting factors of this method is the measurement of the cross sections ratio of the reactions $e^+e^- \rightarrow B^{0} \overline{B}$$^{*0}, B^{*0} \overline{B}$$^{*0}$ with high accuracy. The statistical error ($B$-mesons should be completely reconstructed) does not allow to determine the value of $f_s$ with necessary accuracy~\cite{Belle:2021lzm}.
 
In the present paper, we suggest the procedure to measure $f_s$ which does not require the knowledge of the branching ratios of $\mathit{\Upsilon}(5S)$ decays. It is based on the measurement of time dependence of events when both $B$-mesons produced in $\mathit{\Upsilon}(5S)$ resonance decay semileptonically to the leptons of the same or opposite signs. This method is based on the assumption of isotopic invariance in $\mathit{\Upsilon}(5S)$ decays, $N(B^0 \overline{B}$$^0) = N(B^+B^-)$~\footnote{In case the isotopic invariance in these decays is substantially violated \cite{Bondar:2022kxv}, this problem can be overcome by direct measurement of the yields ratio of charged and neutral $B$ mesons using the full $B$ meson reconstruction as it was done in \cite{Belle:2021lzm}. The reconstruction efficiency ratio for the charged and neutral $B$ mesons could be calibrated at $\mathit{\Upsilon}(4S)$ \cite{Belle:2002lms}.}.
Our approach allows to determine the relative number of $B_s$, $B^0$ and $B^+$ mesons produced in $\mathit{\Upsilon}(5S)$ decays without model assumptions. Thus it makes possible to determine the ratio $\mathcal{B}(B_s \rightarrow D_s \pi^+)/\mathcal{B}(B^0 \to D^- \pi^+)$ at Belle II and to use $B_s \rightarrow D_s \pi^+$ as a normalization channel in order to extract the value of $\mathcal{B}(B_s \rightarrow \mu^+\mu^-)$ at the LHC. We hope that the obtained  accuracy could be comparable with the theoretical one.

The paper is organized as follows:
in Sec.~2 the necessary formulas are presented, in Sec.~3 the difference of the lifetimes of $B^+$ and $B^0$ mesons is taken into account. In order to derive the formulas in Sec.~2 and 3 the time dependences of the $B\bar B$ wave functions are needed. They are obtained in Appendix B for $C$-odd and $C$-even wave functions. In Appendix C the case of one charged $B$-meson is considered. In this case only one $B$-meson is neutral and oscillates. 
In Sec.~4 we discuss the experimental feasibility of the method. In Sec.~5 we present the applications of our method to $\Upsilon(4S)$ decays. In Sec.~6 we conclude.

\section{Branching of the $\Upsilon(5S) \to B^{(*)}_s\bar B^{(*)}_s$ decay}

We consider the process $e^+e^- \to \Upsilon(5S) \to B\bar B X$ with further semileptonic decays of both $B$ mesons. The possible combinations of $B\bar B X$ are listed in Table 1. At Belle II the dependence of the number of events with the leptons of the same sign $dN_{\pm\pm}/d\Delta t$ and the opposite signs $dN_{\pm\mp}/d\Delta t$ on the time interval between the decays $\Delta t$ can be measured. Experimental fit of this dependence allows to extract the branching of the $\Upsilon(5S)\to B^{(*)}_s\bar B^{(*)}_s$ decays.  

\begin{table}[h]
\caption{\label{tab:decaymodes} $\Upsilon(5S)$ decay modes} 
\begin{tabularx}{1.\textwidth}{@{\extracolsep{\fill}}c | c c c | c}
\hline 
\hline 
$C$-parity of $B\bar B$ pair & \multicolumn{3}{c}{Decay modes} & Branching notation \\
\hline
\multicolumn{4}{c}{\multirow{2}{*}{$B^0 \bar{B}^0$ among the final state particles}}\\
\multicolumn{4}{c}{} \\
\hline
& & & \\
\multirow{2}{*}{$C$-odd state} & $B^0 \bar{B}^0$ & $B^{0*} \bar{B}^{0*}$ & $B^0 \bar{B}^0\pi^0$ & \multirow{2}{*}{$(\epsilon_{00})^{\text{odd}}$}\\
 & $B^{0*} \bar{B}^{0*} \pi^0$ & & &\\
 \multirow{2}{*}{$C$-even state} & $B^0 \bar{B}^{0*}$ & $B^{0*} \bar{B^{0}}$ & $B^{0*} \bar{B}^0\pi^0$ & \multirow{2}{*}{$(\epsilon_{00})^{\text{even}}$}\\
 & $B^{0} \bar{B}^{0*} \pi^0$ & & &\\
 & & & \\
\hline
\multicolumn{4}{c}{\multirow{2}{*}{$B^+ B^-$ among the final state particles}}\\
\multicolumn{4}{c}{} \\
\hline
& & & \\
\multirow{3}{*}{$C$-odd and $C$-even states} & $B^+ B^-$ & $B^{+*}B^{-*}$ & $B^+ B^- \pi^0$ & \multirow{3}{*}{$\epsilon_{+-}$}\\
 & $B^{+*} B^{-*} \pi^0$ & & &\\
 & $B^{+} B^{-*}$ & $B^{+*} B^{-}$
  & $B^{+*} B^{-} \pi^0$ &  \\
  & $B^{+} B^{-*} \pi^0$ & & &\\
  & & & \\
\hline
\multicolumn{4}{c}{\multirow{2}{*}{$B_s \bar{B}_s$ among the final state particles}}\\
\multicolumn{4}{c}{} \\
\hline
& & & \\
$C$-odd state & $B_s \bar{B}_s$ & $B_s^{*} \bar{B}_s^{*}$ & & $(\epsilon_{SS})^{\text{odd}}$\\
$C$-even state & $B_s^* \bar{B}_s$ & $B_s \bar{B}_s^*$ & & $(\epsilon_{SS})^{\text{even}}$\\
 & & & \\
\hline
\multicolumn{4}{c}{\multirow{2}{*}{$B^{\pm}B^0$ among the final state particles}}\\
\multicolumn{4}{c}{} \\
\hline
& & & \\
\multirow{2}{*}{No definite $C$-parity} & $B^+ \bar{B}^0 \pi^-$ & $B^{+*}\bar{B}^0 \pi^-$ & $B^+ \bar{B}^{0*} \pi^-$ \; $B^{+*} \bar{B}^{0*} \pi^-$ & $\epsilon_+$\\
 & $B^{-} B^{0} \pi^+$ & $B^{-*} B^0 \pi^+$ & $B^{-}B^{0*} \pi^+$ \; $B^{-*} B^{0*} \pi^+$ & $\epsilon_-$\\
  & & & \\
\hline
\hline 
\end{tabularx} 
\end{table} 

Let us obtain the time-dependent numbers of events with the leptons in the final state using the formulas derived in the Appendices B and C\footnote{The time of $B^*$ radiative decay being approximately $5$ orders of magnitude smaller than the life time of $B$-meson should be safely neglected (see \cite{BM} for the corresponding calculations and further references). With this accuracy $B$-mesons from $B^*$ decays should be considered as produced right at the point of $\Upsilon(5S)$ decay.}. Using the integrands from \eqref{eq:nppoddfull}, \eqref{eq:nppevenfull} and \eqref{eqq:np}, for $dN_{\pm\pm}/d\Delta t$ we have:
\begin{align}\label{eq:dndtpp}
	&\frac{dN_{++}}{d\Delta t} =  L \sigma_{e^+e^- \rightarrow \Upsilon(5S)}\mathcal{B}^2(B^0 \rightarrow \ell^+ \nu_{\ell} X) \frac{\Gamma}{2} e^{-\Gamma \Delta t}\bigg[(\epsilon_{00})^{\text{odd}} \sin^2 \frac{\Delta m \Delta t}{2} + \\ \nonumber  &+\frac{(\epsilon_{00})^{\text{even}}}{2 (1 + x^2)} \bigg( 1 + x^2 - \cos\Delta m \Delta t + x \sin \Delta m \Delta t\bigg) + \frac{(\epsilon_{SS})^{\text{odd}} + (\epsilon_{SS})^{\text{even}}}{2} + \\ \nonumber
	&+\frac{\epsilon_+}{4 + x^2}\bigg(2 + x^2 - 2 \cos\Delta m \Delta t + x \sin \Delta m \Delta t \bigg)\bigg],
\end{align}
where $L$ is the luminosity of the $e^+e^-$ collider, $\sigma_{e^+e^- \rightarrow \Upsilon(5S)}$ is the cross section of the $\Upsilon(5S)$ production, $\mathcal{B}(B^0\rightarrow \ell^+ \nu_{\ell} X)$ is the branching ratio of the semileptonic $B^0$-meson decay, $\Gamma$ is $B^0$ width and $x=\Delta m/\Gamma$. The analogous expression holds for $dN_{--}/d\Delta t$. The branching ratios $\epsilon$ are defined in Table 1\footnote{The $C$-parity of produced pair of neutral $B$-mesons determines the time-dependences of their semileptonic decays. The necessary formulas are given in Appendices B1 and B2. When only one neutral $B$-meson is produced the time-dependences of its decay are given by formulas in Appendix C.}. The following assumptions were used: $\Gamma_H = \Gamma_L = \Gamma$ (see Appendix A), $\Gamma_{B^0} = \Gamma_{B^+} = \Gamma_{B_s}$\footnote{The fact that $\Gamma_{B^+}$ is by $8\%$ smaller than $\Gamma_{B^0} = \Gamma_{B_s}$ will be taken into account in Sec.3.}. We performed the averaging over the time of the numbers of events for the strange $B$ mesons since their fast oscillations can not be resolved by a detector ($\Delta m_S/\Gamma_S = 27.03 \pm 0.09$ \cite{ParticleDataGroup:2022pth}): $<\sin^2{\Delta m_s \Delta t/2}> = 1/2$, $<\cos{\Delta m_s \Delta t/2}> = <\sin{\Delta m_s \Delta t/2}> = 0$.

The analogous formulas are obtained for the case of dileptons of the opposite signs $dN_{\pm\mp}/d\Delta t$ with the help of equations \eqref{eq:npmoddfull}, \eqref{eq:npmevenfull} and \eqref{eqq:nn}:
\begin{align}\label{eq:dndtpm}
	&\frac{dN_{+-}}{d\Delta t} = \frac{dN_{-+}}{d\Delta t} = L \sigma_{e^+e^- \rightarrow \Upsilon(5S)} \mathcal{B}^2(B^0 \rightarrow \ell^+ \nu_{\ell} X) \frac{\Gamma}{2} e^{-\Gamma \Delta t}\bigg[(\epsilon_{00})^{\text{odd}}\cos^2 \frac{\Delta m \Delta t}{2} + \\ \nonumber &+\frac{(\epsilon_{00})^{\text{even}}}{2 (1 + x^2)} \bigg( 1 + x^2 + \cos\Delta m \Delta t - x \sin \Delta m \Delta t\bigg) + \epsilon_{+-} + \frac{(\epsilon_{SS})^{\text{odd}} + (\epsilon_{SS})^{\text{even}}}{2} + \\ \nonumber
	&+\frac{\epsilon_+ + \epsilon_-}{2(4 + x^2)}\bigg(6 + x^2 + 2 \cos\Delta m \Delta t - x \sin \Delta m \Delta t \bigg)\bigg].
\end{align}
It is convenient to form the following ratios from \eqref{eq:dndtpp} and \eqref{eq:dndtpm}:
\begin{equation}\label{eq:diffratio}
	\frac{d(N_{+-} + N_{-+} - N_{--} - N_{++})/d\Delta t}{d(N_{+-} + N_{-+} + N_{--} + N_{++})/d\Delta t} = C + A \sin (\Delta m \Delta t + \varphi),
\end{equation}
	\begin{equation}\label{eq:ppratio}
	\frac{d(N_{++} + N_{--})/d\Delta t}{d(N_{+-} + N_{-+} + N_{--} + N_{++})/d\Delta t} = C' - \frac{A}{2} \sin (\Delta m \Delta t + \varphi),
\end{equation}
where the parameters $C, C', A$ and $\phi$ are equal to
\begin{equation}\label{eq:C}
C = \epsilon_{+-} + \bigg(\epsilon_+ + \epsilon_-\bigg) \frac{2}{4 + x^2},
\end{equation}

\begin{equation}\label{eq:C'}
C' = \frac{(\epsilon_{00})^{\text{odd}} + (\epsilon_{00})^{\text{even}} + (\epsilon_{SS})^{\text{odd}} + (\epsilon_{SS})^{\text{even}}}{2} + \bigg(\epsilon_+ + \epsilon_-\bigg) \frac{1 + x^2/2}{4 + x^2}.
\end{equation}

\begin{equation}
 A = -\sqrt{\left((\epsilon_{00})^{\text{odd}} + \frac{(\epsilon_{00})^{\text{even}}}{1+x^2} + \frac{2}{4+x^2}\left(\epsilon_+ + \epsilon_-\right)\right)^2 + \left(\frac{x}{1+x^2}(\epsilon_{00})^{\text{even}} + \frac{x}{4+x^2}\left(\epsilon_+ + \epsilon_-\right)\right)^2}
\end{equation}
\begin{equation}
 \phi = -\arcsin{\left[\frac{(\epsilon_{00})^{\text{odd}} + \frac{(\epsilon_{00})^{\text{even}}}{1+x^2} + \frac{2}{4+x^2}\left(\epsilon_{+} + \epsilon_{-}\right)}{-A}\right]}.
\end{equation}
The experimental measurement of the time dependences of the left sides of \eqref{eq:diffratio} and \eqref{eq:ppratio} allows to obtain the branching ratio $\epsilon_{SS} = (\epsilon_{SS})^{\text{odd}} + (\epsilon_{SS})^{\text{even}}$ from \eqref{eq:C} and \eqref{eq:C'}:
\begin{equation}\label{eq:2cprim}
	2C' - C = \epsilon_{SS} + \bigg( \epsilon_+ + \epsilon_- \bigg) \frac{x^2}{4 + x^2},
\end{equation}
where we used the isotopic invariance: $\epsilon_{+-} = (\epsilon_{00})^{\text{odd}} + (\epsilon_{00})^{\text{even}}$. Let us estimate the uncertainty in $\epsilon_{SS}$ determination which comes from the last term. 
The decays  of $\Upsilon(5S)$ to $B\bar B\pi$, $\left(B\bar B^* + B^*\bar B\right)\pi$ and $B^*\bar B^*\pi$ contribute to $\epsilon_+$ and $\epsilon_-$ (see Table 1). According to \cite{pdglive} the decays into $\left(B\bar B^* + B^*\bar B\right)\pi$ dominate and their branching ratio equals $(7\pm 2)\%$. Three types of state contribute to this branching ratio: with the radiation of the $\pi^+$, $\pi^-$ and $\pi^0$ mesons. Since $\left(B\bar B^* + B^*\bar B\right)\pi$ has zero isospin, the branching ratio of the production of $\pi^+$ accompanied by $B^{-*}B^0$ or $B^{-}B^{0*}$ ($\epsilon_-$) is equal to that of $\pi^-$ accompanied by $B^{+*}\bar B^0$ or $B^{+}\bar B^{0*}$ ($\epsilon_+$). And equals the branching ratio of the production of $\pi^0$ accompanied by $B^{-*}B^+$ or $B^-B^{+*}$ or $B^0\bar B^{0*}$ or $B^{0*}\bar B^{0}$. Thus $\epsilon_+ + \epsilon_- = (2/3)\times \left(7\pm 2\right)\%$ and the contribution to Eq.\eqref{eq:2cprim} equals  

\begin{equation}
\frac{2}{3}\times \left(7\pm 2\right)\% \times \frac{\left(0.77\right)^2}{4+\left(0.77\right)^2} = \left(0.62 \pm 0.17\right)\%,
\end{equation}
where $x$=0.77 is substituted. Comparing it with the approximate $\epsilon_{SS}$ value $(22\pm 2)\%$ \cite{Belle:2023yfw} we see that the relative uncertainty due to this correction is about $1\%$. 

The approach developed faces  difficulty connected with the distance between B-mesons in
the longitudinal direction before the first of them decays. This distance is negligible
in the case of $\Upsilon(4S)$ decay because the velocities of the produced B-mesons in the center of mass system are negligible compared to their velocity in the laboratory system. This is due to the small energy release in the decay of $\Upsilon(4S)$. However, in the case of $\Upsilon(5S)$ the energy release is not that small, and for zero transverse momentum
the separation of B-mesons prior to decays is comparable to the separation  due to the difference of the decay moments. And in the experiment in the case of semileptonic decays only the sum of these two distances in longitudinal direction is measured.

To resolve this issue one can demand larger velocity of $\Upsilon(5S)$ in the lab system.
Relativistic effects diminish the difference of the longitudinal velocities of the
produced B-mesons and the velocities of $\Upsilon(5S)$ $V=0.6c-0.7c$ do the job (let us remind that now at SuperKEKB $V\approx 0.3c)$. 
In order to clarify this statement let us consider the worst case: $B$ mesons produced in $\Upsilon(5S)$ decays  have zero transverse momenta. In numerical estimates we put $M_{\Upsilon(5S)} = 10.89~GeV$, $M_B = 5.28~GeV$. The velocities of $B$ mesons in the $\Upsilon(5S)$ rest frame are equal to $v=0.24c$. If the velocity of $\Upsilon(5S)$ in the lab frame equals $V=0.6c$ then the velocity of the fast $B$ meson $v_1 = 0.73c$ while that of the slow one is $v_2 = 0.42c$. Let us suppose that the first semileptonic decay is that of the slow $B$-meson and it happens at the time $t = \tau \equiv 1/\Gamma$ in its rest frame. At this moment the separation between fast and slow $B$-mesons is $(v_1 - v_2)\gamma_2\tau$, where $\gamma_2 = (1-v^2_2)^{-1/2}$. We assume further that the fast $B$-meson decays after the time $\tau \equiv 1/\Gamma$ passes in its rest frame after the first decay. Thus the distance between two decay points gets an extra term $l_0 = v_1 \gamma_1\tau$. It is the distance which enters our formulas (or better to say the corresponding time). However, the distance between two points of the semileptonic decays,  measured in the experiment, equals $l = \left[\gamma_2\left(v_1 - v_2\right) + \gamma_1 v_1\right]\tau$. So we obtain:
\begin{equation}
    \frac{l}{l_0} = 1 + \frac{\gamma_2}{\gamma_1}\left(1 - \frac{v_2}{v_1}\right) = 1.3.
\end{equation}
For $V=0.7c$ this ratio equals $1.2$ that is not far from one. In the ultrarelativistic limit $V \to c$ we get $l/l_0 = 1$, while in the nonrelativistic limit $v_2$ becomes negative and $l/l_0 = 3$.
Although the facilities which produce $\Upsilon(5S)$ mesons with large velocity do not exist, their construction in the future is not excluded. They would allow to study time-dependencies of decays of B mesons produced by $\Upsilon(5S)$. 

Another option is to restore momenta
of decaying semileptonically B-mesons like it was done by BaBar collaboration \cite{BaBar:momenta}
and to take into account only the events with the B-meson transverse momenta close to maximum.
Their separation in the longitudinal direction at the moment of the decay of the first
B-meson is negligible in comparison to the distance in the longitudinal direction which the second one flies before it decays.

\section{Corrections due to the longer lifetime of $B^+$ meson}

While the lifetimes of $B^0$ and $B_s$ mesons coincide with the percent accuracy, the lifetime of $B^+$ meson is considerably higher: $\tau_{B^+}/\tau_{B^0} = 1.076\pm 0.004$ \cite{ParticleDataGroup:2022pth}. In this section we will take into account the difference between $B^+$ and $B^0(B_s)$ lifetimes.

The following notation will be used: $\Gamma= \Gamma_{B^0} = \Gamma_{B_s} = \Gamma_{B^+} + \delta$, where $\delta$  is $8\%$ of $\Gamma$ numerically. There are two types of $\delta$ dependences in our formulas: the power and exponential ones. Since in the exponential terms $\delta$ is multiplied by the time interval $\Delta t$ we do not expand it over $\delta$. We expand the power terms keeping corrections of the order of $\delta$ and take into account the exponential terms explicitly since at large time intervals $\Delta t \sim (4 - 5)/\Gamma$ the term of the order of $\delta^2$ becomes essential producing the correction of the order of few percents.

For the terms entering Eqs.\eqref{eq:diffratio} and \eqref{eq:ppratio} we get:
\begin{align}\label{11}
    &\frac{dN_{++}}{d\Delta t} + \frac{dN_{--}}{d\Delta t} + \frac{dN_{+-}}{d\Delta t} + \frac{dN_{-+}}{d\Delta t} = L \sigma_{e^+e^- \rightarrow \Upsilon(5S)} \mathcal{B}^2(\text{B}^0 \rightarrow \ell^+ \nu_{\ell} X)\Gamma e^{-\Gamma \Delta t}\times \\ \nonumber
    &\times \Bigg[ 1 + \frac{\epsilon_+ + \epsilon_-}{2}\left(e^{\delta\Delta t} - 1 + \frac{\delta}{2\Gamma}\left(1+e^{\delta\Delta t}\right)\right) +\epsilon_{+-}\left(e^{\delta\Delta t} - 1 + \frac{\delta}{\Gamma}e^{\delta\Delta t}\right)
  \Bigg],
\end{align}
\begin{equation}\label{12}
\frac{d(N_{+-} + N_{-+} - N_{--} - N_{++})/d\Delta t}{d(N_{+-} + N_{-+} + N_{--} + N_{++})/d\Delta t} = \tilde{C} + \tilde{A} \sin (\Delta m \Delta t + \tilde{\varphi}),
\end{equation}
\begin{equation}\label{13}
\frac{d(N_{++} + N_{--})/d\Delta t}{d(N_{+-} + N_{-+} + N_{--} + N_{++})/d\Delta t} = \tilde{C}' - \frac{\tilde{A}}{2} \sin (\Delta m \Delta t + \tilde{\varphi}),
\end{equation}
where the parameters $\tilde C$, $\tilde C'$, $\tilde A$ and $\tilde \varphi$ equal
\begin{equation}\label{14}
    \tilde C = \frac{\frac{\epsilon_{+} + \epsilon_{-}}{4+x^2}e^{\delta\Delta t}\left(2+\frac{4-x^2}{4+x^2}\frac{\delta}{\Gamma}\right) + \epsilon_{+-}e^{\delta\Delta t}\left(1+\frac{\delta}{\Gamma}\right)}{ 1 + \frac{\epsilon_+ + \epsilon_-}{2}\left(e^{\delta\Delta t} - 1 + \frac{\delta}{2\Gamma}\left(1+e^{\delta\Delta t}\right)\right) + \epsilon_{+-}\left(e^{\delta\Delta t} - 1 + \frac{\delta}{\Gamma}e^{\delta\Delta t}\right)},
\end{equation}
\begin{equation}
    \tilde C' = \frac{\frac{\epsilon_{00}}{2} + \frac{\epsilon_{SS}}{2} + \frac{\epsilon_+ + \epsilon_-}{4}\left( 1 + e^{\delta\Delta t}\frac{x^2}{4+x^2} + \frac{\delta}{2\Gamma}\left(1 + e^{\delta\Delta t}\frac{x^2\left(12+x^2\right)}{\left(4+x^2\right)^2} \right) \right)}{1 + \frac{\epsilon_+ + \epsilon_-}{2}\left(e^{\delta\Delta t} - 1 + \frac{\delta}{2\Gamma}\left(1+e^{\delta\Delta t}\right)\right) + \epsilon_{+-}\left(e^{\delta\Delta t} - 1 + \frac{\delta}{\Gamma}e^{\delta\Delta t}\right)},
\end{equation}
\begin{align}
    \tilde A = -\frac{\sqrt{a^2 + b^2}}{1 + \frac{\epsilon_+ + \epsilon_-}{2}\left(e^{\delta\Delta t} - 1 + \frac{\delta}{2\Gamma}\left(1+e^{\delta\Delta t}\right)\right) + \epsilon_{+-}\left(e^{\delta\Delta t} - 1 + \frac{\delta}{\Gamma}e^{\delta\Delta t}\right)},
\end{align}
where 
\begin{align}
    &a = (\epsilon_{00})^{\text{odd}} + \frac{(\epsilon_{00})^{\text{even}}}{1+x^2} + \frac{\epsilon_+ + \epsilon_-}{4+x^2}\left(2 + \frac{4-x^2}{4+x^2}\frac{\delta}{\Gamma}\right),\\ \nonumber
    &b = \frac{x}{1+x^2}(\epsilon_{00})^{\text{even}} + \left(\epsilon_+ + \epsilon_-\right)\frac{x}{4+x^2}\left(1 + \frac{4}{4+x^2}\frac{\delta}{\Gamma}\right),
\end{align}
\begin{equation}
    \tilde \varphi = -\arcsin{\left[\frac{(\epsilon_{00})^{\text{odd}} + \frac{(\epsilon_{00})^{\text{even}}}{1+x^2} + \frac{\epsilon_{+} + \epsilon_{-}}{4+x^2}\left(2 + \frac{4-x^2}{4+x^2}\frac{\delta}{\Gamma}\right)}{-\tilde{A}\left(1 + \frac{\epsilon_+ + \epsilon_-}{2}\left(e^{\delta\Delta t} - 1 + \frac{\delta}{2\Gamma}\left(1+e^{\delta\Delta t}\right)\right) + \epsilon_{+-}\left(e^{\delta\Delta t} - 1 + \frac{\delta}{\Gamma}e^{\delta\Delta t}\right)\right)}\right]}.
\end{equation}
Finally Eq.\eqref{eq:2cprim} is substituted by:
\begin{align}\label{18}
    &2\tilde C' - \tilde C = \\ \nonumber
    &=\frac{\epsilon_{00} + \epsilon_{SS} - \epsilon_{+-}e^{\delta\Delta t}\left(1+\frac{\delta}{\Gamma}\right) + \frac{\epsilon_+ + \epsilon_-}{4+x^2}\left(2 + \frac{x^2}{2} - e^{\delta\Delta t}\left(2-\frac{x^2}{2}\right) + \frac{\delta}{\Gamma}\left(1 + \frac{x^2}{4} + e^{\delta\Delta t}\frac{x^4 + 16x^2-16}{4\left(4+x^2\right)}\right)\right)}{1 + \frac{\epsilon_+ + \epsilon_-}{2}\left(e^{\delta\Delta t} - 1 + \frac{\delta}{2\Gamma}\left(1+e^{\delta\Delta t}\right)\right) + \epsilon_{+-}\left(e^{\delta\Delta t} - 1 + \frac{\delta}{\Gamma}e^{\delta\Delta t}\right)}.
\end{align}
Let us note that the terms proportional to $(\epsilon_{+} +  \epsilon_{-})/2$ in \eqref{11}
and in the denominators of \eqref{14} - \eqref{18} are less than one percent for $\Delta t \leq 3/\Gamma$ and can be safely omitted. 

Taking into account that $\epsilon_{00} + \epsilon_{SS} +\epsilon_{+-}
+ \epsilon_+ + \epsilon_- = 1$ by construction, one can extract the value of $\epsilon_{+-}$ from the combination $ 2\tilde C' - \tilde C$ 
and with the help of the isotopic symmetry  ($\epsilon_{00} =
\epsilon_{+-}$) we get $\epsilon_{SS}$. The correction due to the isotopic symmetry violation may be accounted for as well.

\begin{figure}
\begin{subfigure}{0.5\textwidth}
    \includegraphics[width=\textwidth]{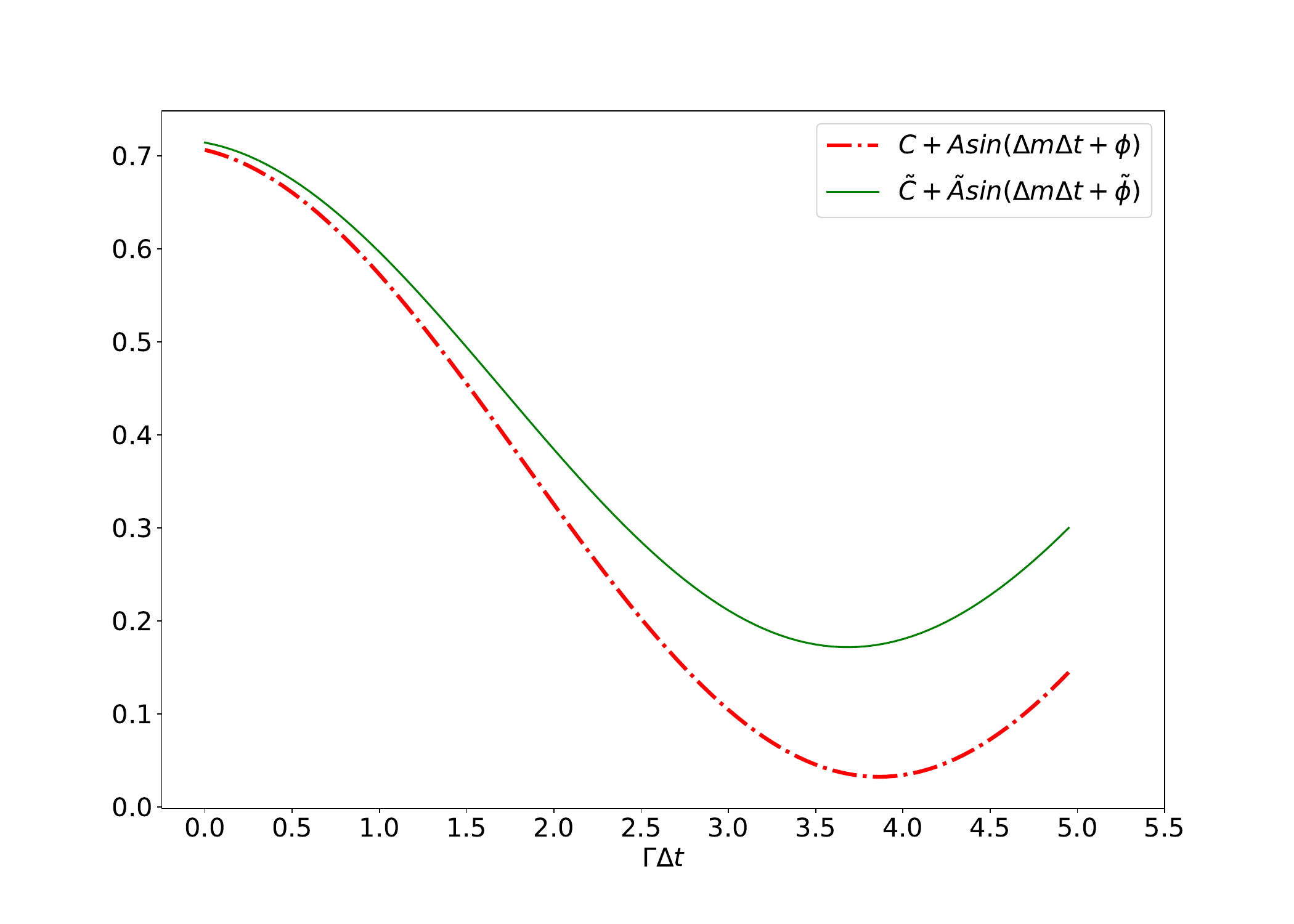}
    \label{fig1}
    \end{subfigure}
    \begin{subfigure}{0.5\textwidth}
    \includegraphics[width=\textwidth]{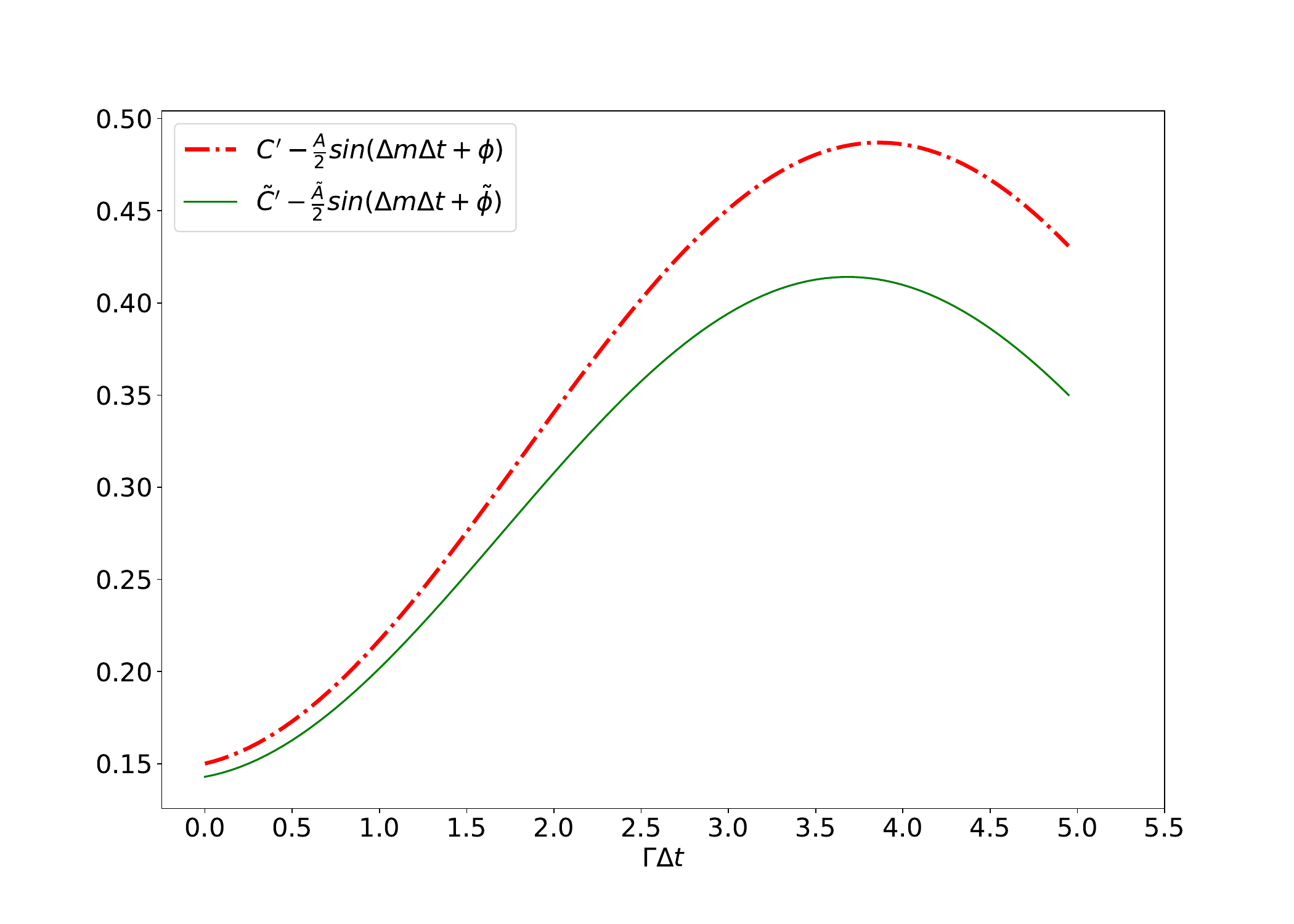}
    \label{fig2}
    \end{subfigure}
    \caption{The functions \eqref{eq:diffratio}, \eqref{12} and \eqref{eq:ppratio}, \eqref{13}. 
    \label{fig}
    }
\end{figure}

Functions \eqref{eq:diffratio}, \eqref{12} and \eqref{eq:ppratio}, \eqref{13} are shown in Fig.\ref{fig} correspondingly for the experimental values of the branching ratios $\epsilon$ \cite{Belle:2021lzm} which approximately equal: $\epsilon_{SS}=20\%$, $\epsilon_{+-}=38\%$, $\left(\epsilon_{00}\right)^{\text{odd}}=27\%$, $\left(\epsilon_{00}\right)^{\text{even}} = 11\%$ and $\epsilon_{\pm}=2.5\%$.  

\section{Experimental feasibility of the method and assessment of statistical accuracy}

It should be noted that the existing experience of experiments at $B$ factories shows that semileptonic decays of $B$ mesons are reliably distinguished from possible backgrounds even for a fully inclusive method of the event registration~\cite{BaBar:2016rxh,BaBar:2001bcs}~\footnote{Based on the study of the inclusive lepton spectrum in~\cite{BaBar:2016rxh}, it can be concluded that the background level from direct charm production in $e^+e^-$ annihilation is small, contributes only to events with opposite lepton charges, and can be effectively taken into account based on data obtained below the B meson production threshold.}. At a sufficiently high threshold on the momentum of the prompt lepton ($b \to cl\nu$), the contribution of events with successive leptons ($b \to c \to l\nu$) becomes insignificant. This is confirmed by successful measurements of $\Delta m_d$ in events with doubly semileptonic decays of $B^0 \bar{B^0}$ on $\Upsilon(4S)$~\cite{BaBar:2001bcs}. As for the expected statistical accuracy of the method, it can be estimated by following the estimate made in~\cite{Sia:2006cq} for the integrated luminosity of $\Upsilon(5S)$ $120~\text{fb}^{-1}$, which has already been used in many measurements of the Belle detector, for example~\cite{Belle:2021lzm}. We can estimate the error in $f_s$ by taking the $C$-odd contribution as $70\%$, the $C$-even component as $30\%$, no incoherent $B^0\bar{B^0}$ contribution and $f_s$ equal to 20$\%$ from the recent Belle measurements~\cite{Belle:2023yfw}. Taking into account the semileptonic branching ratio ($10.5\%$), the fraction of high momentum leptons ($p_{min} = 1.3~\text{GeV}$) above the minimum lepton momentum cut ($0.6$), and the lepton efficiency ($0.6$), we estimate that the fractional error of $\pm 2\%$ on $f_s$ can be achieved with $120~\text{fb}^{-1}$ of data. At present, as noted earlier, the error in the value of $f_s$ is of the order of $10\%$~\cite{Belle:2023yfw}. Of course, additional data on $\Upsilon(5S)$, which the Belle-II detector is expected to be able to obtain, will further reduce the statistical uncertainty. As for the estimation of systematic uncertainties of the method, this requires realistic modelling of both the detector parameters and different backgrounds, which will probably be possible to do in a real experiment.

\section{Applications to the $\Upsilon(4S)$ decays}

 The branching ratios $\mathcal{B}\left(\Upsilon(4S)\to B^0 \bar B^0\right) \equiv \epsilon_{00}$ and $\mathcal{B}\left(\Upsilon(4S)\to B^+ B^-\right) \equiv \epsilon_{+-}$  can be found from the time dependences of semileptonic decays of both $B$-mesons produced in the $\Upsilon(4S)$ decay. Taking into account that 
\begin{equation}
    \frac{dN_{++}}{d\Delta t} = \frac{dN_{--}}{d\Delta t} \sim \epsilon_{00}\sin^2{\frac{\Delta m\Delta t}{2}},
\end{equation}
\begin{equation}
    \frac{dN_{+-}}{d\Delta t} = \frac{dN_{-+}}{d\Delta t} \sim \epsilon_{00}\cos^2{\frac{\Delta m\Delta t}{2}} +   \epsilon_{+-},
\end{equation}
we get:
\begin{equation}\label{23}
\frac{d\left(N_{+-}  + N_{-+} - N_{++} - N_{--}\right)/d\Delta t}{d\left(N_{+-}  + N_{-+} + N_{++} + N_{--}\right)/d\Delta t} = D + B\cos{\Delta m\Delta t},
\end{equation}
\begin{equation}\label{24}
\frac{d\left(N_{++}  + N_{--}\right)/d\Delta t}{d\left(N_{+-}  + N_{-+} + N_{++} + N_{--}\right)/d\Delta t} = \frac{1}{2}B\left(1 -\cos{\Delta m\Delta t}\right),
\end{equation}
where $D = \epsilon_{+-}$ and $B = \epsilon_{00}$.

The experimental measurement of the time dependences of the left sides of  \eqref{23} and \eqref{24} allows to obtain the branching ratios $\epsilon_{+-}$ and $\epsilon_{00}$. Current result is $\epsilon_{+-}/\epsilon_{00} = 1.058 \pm 0.024$ \cite{pdglive}. We hope that our approach allows to reduce current uncertainties which is important to more accurately measure the probabilities of all exclusive $B^+$, $B^0$ decays in future Belle-II experiments. The corrections due to the longer lifetime of $B^+$ should be accounted for in the same way as it is done in Sec.3.

Simultaneous decays of both $B$-mesons into dileptons of the same sign may occur due to the contribution of $\Upsilon(4S) \rightarrow B^0 \bar B^0 \gamma$ decay. It's relative probability appeared to be very small,  $\mathcal{B}(\Upsilon(4S) \rightarrow B^0 \bar B^0 \gamma) \approx 3\times 10^{-9} $ according to calculations made in \cite{colangelo}. In order to obtain experimental bound one should look for decays of $bb$-quarks (or $\bar b \bar b$-quarks) containing $B$-mesons in the limit $\Delta t\rightarrow 0$.

As a one more example of the application of two-particle wave functions, let us consider the decay $\mathit{\Upsilon}(4S) \rightarrow B^0\overline{B}$$^0 \rightarrow J/\psi K_S J/\psi K_S$. For the amplitude of this decay we have
\begin{eqnarray}
	\langle J/\psi K_S, J/\psi K_S|\psi(t_1, t_2)\rangle_{\text{odd}} = e^{-2i M t - \Gamma t} \bigg[ A \cos\bigg(\frac{\Delta m}{2}t_1\bigg) + i \frac{q}{p}\overline{A}\sin\bigg(\frac{\Delta m }{2}t_1\bigg)\bigg]\times \nonumber\\ \times \bigg[\overline{A}\cos\bigg(\frac{\Delta m}{2}t_2 \bigg) + i\frac{p}{q}A\sin\bigg(\frac{\Delta m}{2}t_2\bigg)\bigg] - (t_1 \longleftrightarrow t_2) = e^{-2 i M t - \Gamma t} \bigg[ \bigg(i \frac{p}{q}A^2 - \nonumber\\ - i \frac{q}{p}\overline{A}^2\bigg) \cos\bigg( \frac{\Delta m}{2}t_1\bigg) \sin\bigg( \frac{\Delta m}{2}t_2 \bigg) + \bigg( i \frac{q}{p}\overline{A}^2  - i \frac{p}{q} A^2 \bigg)\sin\bigg(\frac{\Delta m}{2}t_1\bigg) \cos\bigg( \frac{\Delta m}{2}t_2\bigg) \bigg]  \nonumber\\ = -e^{-2i M t - \Gamma t} \bigg( i \frac{p}{q} A^2 \bigg) \bigg[1 - \lambda^2 \bigg] \sin\bigg(\frac{\Delta m \Delta t}{2}\bigg),
\end{eqnarray}
where $A$($\bar A$) is the amplitude of $B^0(\bar B^0) \to J/\psi K_S$ decay and $\lambda = (q/p)(\overline{A}/A)$ was introduced. In the last line the fact that $A=\bar A$ was used. The ratio $q/p$ can be expressed via the angle $\beta$ of the unitarity triangle: $q/p = e^{-2i\beta}$. Then the probability of the decay equals
\begin{eqnarray}\label{pjpsi}
	\mathbb{P}(J/\psi K_S, J/\psi K_S) = e^{-2 \Gamma t} A^4 [1 - e^{4 i \beta}][1 - e^{-4 i \beta}] \sin^2\bigg( \frac{\Delta m \Delta t}{2} \bigg) = \nonumber \\ = e^{-2 \Gamma t} A^4 \cdot 4 \sin^2(2 \beta) \sin^2 \bigg( \frac{\Delta m \Delta t}{2} \bigg).
\end{eqnarray}
It equals zero for $\Delta t=0$ due to Bose statistics, for $\Delta m=0$ because of the absence of oscillations and for $\beta=0$ due to $CP$ conservation. The point is that $\mathit{\Upsilon}(4S)$ is $CP$ even state while the final state is $CP$ odd.

Changing integration variables to $t$ and $\Delta t$ according to (\ref{eq:measure}), integrating over the phase space of produced particles and dividing by the $B$-mesons widths for the decay branching we get:

\begin{align}\label{branch}
	\mathcal{B}(\mathit{\Upsilon}(4S) \rightarrow J/\psi K_S J/\psi K_S) &= 8 \cdot 0.5 \cdot\mathcal{B}^2(B^0 \to J/\psi K_S)\Gamma^2 \sin^22\beta \times \\ \nonumber \times\int\limits_{0}^{+\infty}d\Delta t\int\limits_{\Delta t/2}^{+\infty}dt e^{-2 \Gamma t} \sin^2 \bigg(\frac{\Delta m \Delta t}{2} \bigg) 
	\label{eq:NJpsi} 
	&=0.5 \cdot 2\sin^22\beta \bigg( \frac{x^2}{1 + x^2} \bigg) \mathcal{B}^2(B^0 \to J/\psi K_S),
\end{align}
where the factor $\Gamma^2$ takes into account wave function normalization on $1/\Gamma^2$ decaying $B$-meson pairs (see the remark after equation \eqref{eq:npmoddfull}). Factor $0.5$ takes into account branching of $\mathit{\Upsilon}(4S) \to B^0 \bar B^0$.

Let us estimate the number of $\mathit{\Upsilon}(4S) \rightarrow J/\psi K_S J/\psi K_S$ decays which could be detected at Belle II.  The total number of  $\mathit{\Upsilon}(4S)$ which should
be produced at SuperKEKB with integrated luminosity $L=50~\text{ab}^{-1}$ equals $\approx 6 \cdot 10^{10}$. For the number of decays $\mathit{\Upsilon}(4S) \rightarrow J/\psi K_S J/\psi K_S$ with the help of \eqref{branch} we get:
\begin{equation}
    N\left(\mathit{\Upsilon}(4S) \rightarrow J/\psi K_S J/\psi K_S\right) =\epsilon\cdot N\left(\mathit{\Upsilon}(4S)\right)\cdot\mathcal{B}\left(\mathit{\Upsilon}(4S) \rightarrow J/\psi K_S J/\psi K_S\right) \approx 60,
\end{equation}
where $\epsilon = 0.02$ is an efficiency of Belle II detector in the $\mathit{\Upsilon}(4S) \rightarrow J/\psi K_S J/\psi K_S$ channel\footnote{When evaluating the efficiency, it is assumed that one $J/\psi$ is registered by lepton decay, and the second is determined by the recoil mass to $J/\psi K_S K_S$. We are grateful to P.N.Pakhlov for the discussion of the efficiency of the detector Belle II in the considered mode.}. Taking into account the excited charmonium states this number can be doubled. 

Making analogous calculations we can obtain the results for $N(J/\psi K_S J/\psi K_S)$ for $C$-even wave function and for the $B^0\bar B^0$ pair in the incoherent state:

\begin{eqnarray}\label{neven}
	N_{\text{even}}(J/\psi K_S J/\psi K_S) = \frac{A^4}{\Gamma^2}\frac{4 + 5 x^2 + 3 x^4 + 3 x^2\cos4\beta + x^4 \cos4\beta}{(1 + x^2)^2},\\ \label{ninc}
	N_{\text{inc}}(J/\psi K_S J/\psi K_S) = \frac{A^4}{2\Gamma^2} \frac{2 + 3 x^2 + 2 x^4 + x^2\cos4\beta}{(1 + x^2)^2},
\end{eqnarray}
and for the number $N(J/\psi K_S J/\psi K_S)$ for $C$-odd state from \eqref{pjpsi}:
\begin{equation}\label{nodd}
    N_{odd}(J/\psi K_S J/\psi K_S) = \frac{A^4}{\Gamma^2}\frac{x^2\left(1-\cos{4\beta}\right)}{1+x^2}.
\end{equation}
Comparing the equations \eqref{neven}, \eqref{ninc} and \eqref{nodd} we get:
\begin{equation}\label{relation}
    N_{even} + N_{odd} = 4N_{inc}.
\end{equation}
The origin of this equation is the relation between equations \eqref{func:odd}, \eqref{func:even} and \eqref{func:inc}: \\$|\langle J/\psi K_S, J/\psi K_S|\psi(t_1, t_2)\rangle_{\text{odd}}|^2 + |\langle J/\psi K_S, J/\psi K_S|\psi(t_1, t_2)\rangle_{\text{even}}|^2 = 4|\langle J/\psi K_S, J/\psi K_S|\psi(t_1, t_2)\rangle_{\text{inc}}|^2$. 

The decay of the $B^0\bar B^0$ pair in even and incoherent states into $J/\psi K_S J/\psi K_S$ do not violate $CP$ and can occur without $B^0$ - $\bar B^0$ oscillations.

\section{Conclusion}
It was demonstrated that consideration of time dependence of semileptonic decays of both $B$ mesons produced at $\Upsilon(5S)$ resonance allows to determine the fraction of $\Upsilon(5S) \to B^{(*)}_s\bar B^{(*)}_s$ decays with a few percent accuracy. This value can be used to determine with the same accuracy the $\mathcal{B}(B_s \to D_s \pi^+)$ at $\Upsilon(5S)$ by the following relation:
\begin{equation}
\frac{N(B_s \to D_s \pi^+)}{N(B^0 \to D^- \pi^+)}\frac{N(B^0)}{N(B_s)} \mathcal{B}(B^0 \to D^- \pi^+) = \mathcal{B}(B_s \to D_s \pi^+).\footnote{In addition to the uncertainty in $f_s/f_d$ at $\Upsilon(5S)$, the uncertainty in the branching of $\mathcal{B}(B_s \to D_s \pi^+)$ will be determined by the errors in measurement of the decay probabilities of $B^0 \to D^- \pi^+$ and $D \to K^+\pi^-\pi^-$, $D_s \to K^+K^-\pi^+$. At present, the decay probability of $B^0 \to D^-\pi^+$ is known with an accuracy of about $3\%$. The main contribution to the error in this probability is the uncertainty in the yield ratio of neutral and charged $B$ mesons at the maximum of the $\Upsilon(4S)$ cross section $(f_d/f_u)$, which is about $2\%$\cite{Belle:2021udv}. We believe that a method similar to that considered in the paper for determining the $f_s/f_d$ ratio will make it possible to significantly reduce the systematic error in the $f_d/f_u$ ratio on $\Upsilon(4S)$ in future Belle II measurements, see the first part of Sec.5.}
\end{equation}
Thus we suggest the way to determine the probability of the rare decay $B_s \to \mu^+\mu^-$ with a few percent accuracy at the LHC using $B_s \to D_s \pi^+$ as a normalization channel.


\vspace{0.5cm}
\section*{Acknowledgements} 
We are grateful to R.V. Mizuk for carefully reading the manuscript and for useful discussions and to P.N Pakhlov for the valuable discussion. We are also grateful to S.I. Godunov for the help in preparing the manuscript. 

\appendix
\counterwithin*{equation}{section} 
\renewcommand\theequation{\thesection\arabic{equation}} 

\section{$B^0$-meson oscillations}
	

\begin{figure}[H]
		\center{\includegraphics[width = 0.4 \linewidth]{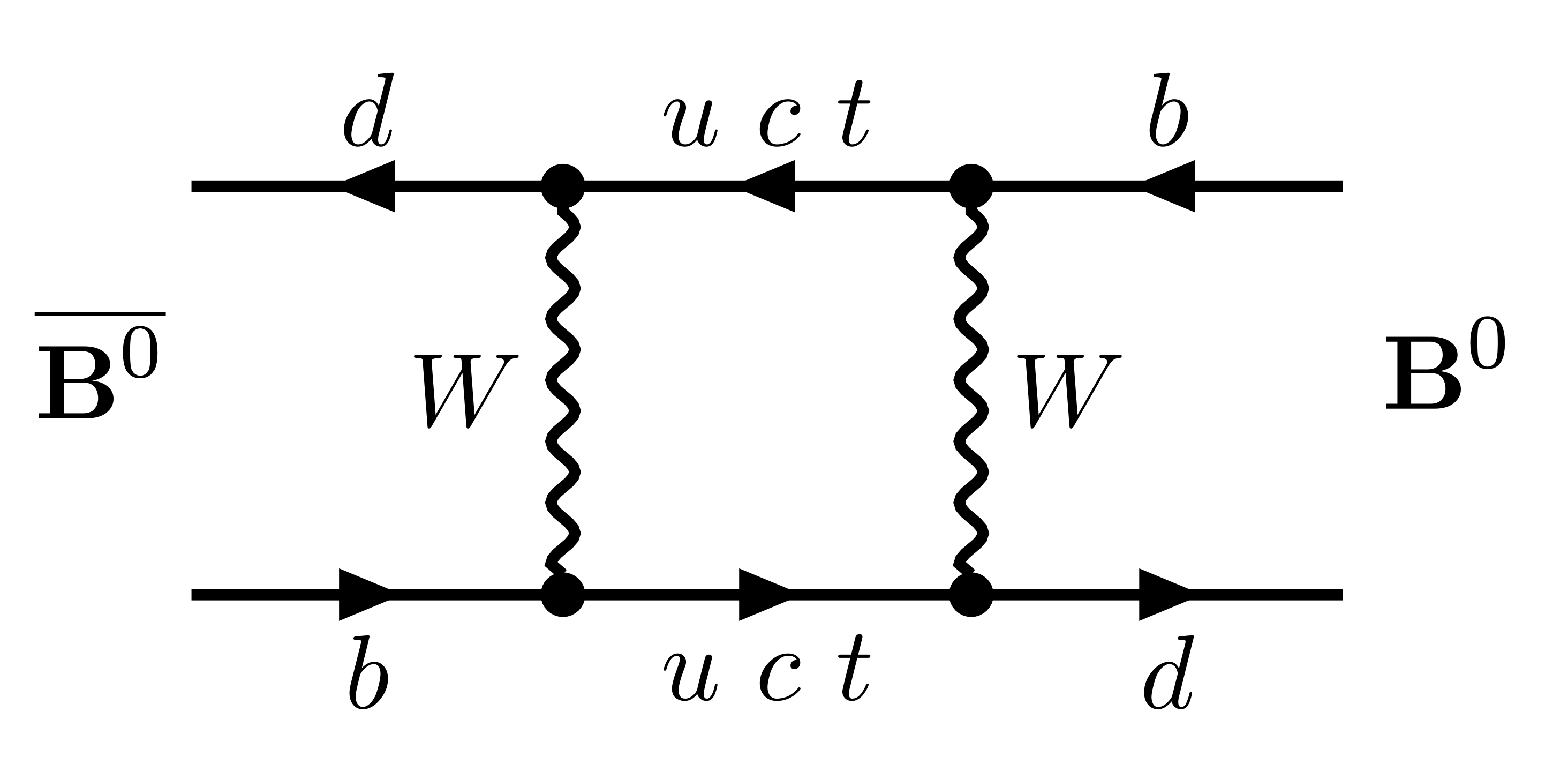}}
		\caption{Feynman diagram describing $B^0-\overline{B}$$^0$ oscillations.} 
		\label{fig:1}
\end{figure} 

The neutral $B^0$-mesons do oscillate, so they are not the states with a definite masses. $B^0$ and $\overline{B}$$^0$ propagate as linear combinations of the states $B_{\text{L}}$ and $B_{\text{H}}$, which are the eigenstates of the Hamiltonian of the system $\hat{\mathcal{H}}$:
\begin{equation}\label{def:mas}
	\begin{gathered} 
        B_{\text{L}} = p B^0 + q \overline{B}\text{}^0, \\ 
        B_{\text{H}} = p B^0 - q \overline{B}\text{}^0, \\ 
      \end{gathered}
\end{equation}
where $p$ and $q$ are the complex parameters describing the mixing of $B^0$-mesons.

Using the definition of the states (\ref{def:mas}), we obtain the expressions for the time evolution of neutral $B$-mesons, $B^0$ and $\overline{B}$$^0 $:
\begin{equation}\label{def:ev}
	\begin{gathered}
		|B^0 (t) \rangle = \frac{e^{- i \lambda_{\text{L}} t} + e^{- i\lambda_{\text{H}} t}}{2} |B^0\rangle + \frac{q}{p} \frac{e^{- i \lambda_{\text{L}} t} - e^{- i\lambda_{\text{H}} t}}{2} |\overline{B}\text{}^0\rangle, \\
		|\overline{B}\text{}^0 (t) \rangle = \frac{e^{- i \lambda_{\text{L}} t} + e^{- i\lambda_{\text{H}} t}}{2} |\overline{B}\text{}^0\rangle + \frac{p}{q} \frac{e^{- i \lambda_{\text{L}} t} - e^{- i\lambda_{\text{H}} t}}{2} |B^0\rangle, \\
	\end{gathered}
\end{equation} 
where $\lambda_{\text{L}}$ and $\lambda_{\text{H}}$ are the eigenvalues of the Hamiltonian
\begin{equation}
	\hat{\mathcal{H}} |B_{\text{L}, \text{H}}\rangle = \lambda_{\text{L}, \text{H}} |B_{\text{L}, \text{H}}\rangle \rightarrow \lambda_{\text{L}, \text{H}} = m_{\text{L}, \text{H}} - i \Gamma_{\text{L}, \text{H}}/2.
\end{equation}
According to \cite{ParticleDataGroup:2022pth} the ratio $(\Gamma_{\text{H}} - \Gamma_{\text{L}})/(\Gamma_{\text{H}} + \Gamma_{\text{L}} )$ is numerically negligible, so we will substitute $\Gamma_{\text{H}} = \Gamma_{\text{L}} \equiv \Gamma$ in our formulas. Let us rewrite (\ref{def:ev}) extracting the common phase factor:
\begin{equation}\label{eq:Bosc}
	\begin{gathered}
		|B^0(t)\rangle = e^{- i M t} e^{-(\Gamma/2)t} \bigg[ \cos\bigg(\frac{\Delta m}{2}t\bigg) |B^0\rangle + i \frac{q}{p} \sin \bigg(\frac{\Delta m}{2} t\bigg) |\overline{B}\text{}^0\rangle \bigg], \\
		|\overline{B}\text{}^0(t)\rangle = e^{- i M t} e^{-(\Gamma/2)t} \bigg[ \cos\bigg(\frac{\Delta m}{2}t\bigg) |\overline{B}\text{}^0\rangle + i \frac{p}{q} \sin \bigg(\frac{\Delta m}{2} t\bigg) |B^0\rangle \bigg],
	\end{gathered}
\end{equation}
where $M = (m_{\text{L}} + m_{\text{H}})/2$ and $\Delta m = m_{\text{H}} - m_{\text{L}}$. Since the modulus of the ratio $|p/q| = 1 + \mathcal{O}(10^{-3})$ \cite{ParticleDataGroup:2022pth}, then we will assume $|p/q| = 1$.

The pair $B^0 \overline{B}$$^0$ produced in the strong decay of $\mathit{\Upsilon}(4S)$ has the quantum numbers $J^{PC} = 1^{--}$. That is, $B^0$ and $\overline{B}$$^0$ are in the $p$-wave. Therefore, the wave function of the pair is antisymmetric with respect to the permutation of $B^0$ and $\overline{B}$$^0$. The semileptonic decay of one $B$-meson tags the remaining one. If $B^0 \rightarrow \ell^+ \nu_{\ell} X$ decay occurs, then another meson at this moment is $\overline{B}$$^0$, and if $\ell^-$ is in the final state of the tagging decay, then the other meson is $B^0$. The remaining $\overline{B}$$^0 (B^0)$-meson starts to oscillate and as a result a lepton of any sign can be produced in its semileptonic decay.

The ratio of the number of dileptons of the same sign to the number of dileptons of opposite signs $R_{\text{odd}}$ can be calculated using single-particle wave functions (\ref{eq:Bosc}). It is given by the formula (\ref{eq:rodd}). The measurement of an unexpectedly large value of this ratio by the ARGUS collaboration \cite{ARGUS:1987xtv} indicated a large value of the $t$-quark mass, since the contribution of the $t$-quark loop to the diagram in Fig.\ref{fig:1} is proportional to $m_t^2$.

Using a one-particle wave function is sufficient to calculate $R$ in the case of an antisymmetric state of the $B^0 \overline{B}$$^0$ pair, but in the case of a symmetric state it is necessary to use a two-particle wave function. The result of calculating $R_{\text{even}}$ is given by the formula (\ref{eq:reven}) and is also known in the literature \cite{Sia:2006cq, 14}. Its derivation, as well as the derivation of the expressions for $R_{\text{inc}}$ (\ref{eq:rinc}), should be based on the two-particle wave functions of the system $B^0\overline{B}$$^0$.

\section{Calculations with two-particle wave function}

Let us consider the two-particle wave function characterizing the pair $B^0\overline{B}$$^0$. In this case a wave function can have different $C$-parity: it can be $C$-even, $C$-odd, or have no definite $C$-parity. Such diversity arises in the decays of $\mathit{\Upsilon}$-mesons or in incoherent production of $B$-mesons in $pp$ collisions.

As already mentioned, the $C$-odd wave function is obtained when $\mathit{\Upsilon}(4S)$ decays into a $B^0 \overline{B}$$^0$ pair, since $\mathit{\Upsilon}$ resonance has quantum numbers $J^{PC} = 1^{--}$. $C$-odd wave function looks like:
\begin{equation}\label{func:odd}
	|\psi(t_1, t_2)\rangle_{\text{odd}} = |B^0(t_1)\rangle\otimes |\overline{B}\text{}^0(t_2)\rangle - |\overline{B}\text{}^0(t_1)\rangle\otimes |B^0(t_2)\rangle.
\end{equation}

The $C$-even wave function is obtained in the decay of $\mathit{\Upsilon}(5S)$ into the pair $B^{0*} \overline{B}$$^0$ or $\overline{B}$$^{0*} B^0$ with the subsequent decay of $B^{0*} \rightarrow B^0 \gamma$ or $\overline{B^{0*}} \rightarrow \overline{B}$$^0 \gamma$. In this case, the wave function is even under permutation of $B$-mesons and is equal to
\begin{equation}\label{func:even}
	|\psi(t_1, t_2)\rangle_{\text{even}} = |B^0(t_1)\rangle\otimes |\overline{B}\text{}^0(t_2)\rangle + |\overline{B}\text{}^0(t_1)\rangle\otimes |B^0(t_2)\rangle.
\end{equation} 

The incoherent production of neutral $B$-mesons is described by the following wave function:
\begin{equation}\label{func:inc}
	|\psi(t_1, t_2) \rangle_{\text{inc}} = |B^0(t_1)\rangle\otimes |\overline{B}\text{}^0(t_2)\rangle.
\end{equation}

Let us calculate for these three wave functions the time dependences of the decay probabilities and the values of $R$, which are nonzero due to the oscillations of $B$-mesons.

\subsection{$C$-odd wave function}

Substituting the expressions (\ref{eq:Bosc}) in (\ref{func:odd}), and taking into account the fact that the amplitudes for the decays $B^0 \rightarrow \ell^+ \nu_{\ell} X$ and $\overline{B}$$^0 \rightarrow \ell^- \overline{\nu}_{\ell} \overline{X}$ are equal, we get:
\begin{eqnarray}\label{eq:nppodd}
	N_{++} = N_{--} = \frac{1}{2} \int\limits_{0}^{+\infty} dt_1 \int\limits_{0}^{+\infty} dt_2 e^{-\Gamma (t_1 + t_2)}\bigg[ \cos\bigg( \frac{\Delta m}{2} t_1 \bigg) \sin\bigg( \frac{\Delta m}{2}t_2 \bigg)  -\nonumber \\ - \sin\bigg( \frac{\Delta m}{2}t_1\bigg) \cos\bigg( \frac{\Delta m}{2}t_2 \bigg) \bigg]^2,
\end{eqnarray}
\begin{eqnarray}\label{eq:npmodd}
	N_{+-} = N_{-+} = \frac{1}{2} \int\limits_{0}^{+\infty} dt_1 \int\limits_{0}^{+\infty} dt_2 e^{-\Gamma (t_1 + t_2)} \bigg[ \cos\bigg(\frac{\Delta m}{2}t_1 \bigg) \cos\bigg(\frac{\Delta m}{2}t_2 \bigg) + \nonumber\\ +
	\sin\bigg(\frac{\Delta m}{2}t_2 \bigg) \sin\bigg(\frac{\Delta m}{2}t_1 \bigg) \bigg]^2.
\end{eqnarray}

It is convenient to change integration variables from times $t_1$ and $t_2$ to their linear combinations $t = (t_1 + t_2)/2$ and $\Delta t = t_1 - t_2$. The integral changes as follows
\begin{equation}\label{eq:measure}
	\int\limits_{0}^{+\infty} dt_1 \int\limits_{0}^{+\infty} dt_2 f(t_1, t_2) = \int\limits_{0}^{+\infty} d(\Delta t) \int\limits_{\Delta t/2}^{+\infty} dt \bigg[ f\bigg(t + \frac{\Delta t}{2}, t - \frac{\Delta t}{2}\bigg) + f\bigg(t - \frac{\Delta t}{2}, t + \frac{\Delta t}{2}\bigg) \bigg].
\end{equation}

Integrating over $d t$ we find how the number of produced dileptons of the same and opposite signs depends on the time interval between semileptonic decays of $B$-mesons:
\begin{equation}\label{eq:nppoddfull}
	N_{++} = N_{--} = \frac{1}{2 \Gamma} \int\limits_{0}^{+\infty} d(\Delta t) e^{-\Gamma \Delta t} \sin^2 \frac{\Delta m \Delta t}{2} = \frac{1}{4 \Gamma^2} \frac{x^2}{1 + x^2},
\end{equation}	
\begin{equation}\label{eq:npmoddfull}
	N_{+-} = N_{-+} = \frac{1}{2 \Gamma} \int\limits_{0}^{+\infty} d(\Delta t) e^{-\Gamma \Delta t} \cos^2 \frac{\Delta m \Delta t}{2}  = \frac{1}{4 \Gamma^2} \frac{2 + x^2}{1 + x^2},
\end{equation}
where the dimensionless parameter $x = \Delta m / \Gamma$ was introduced. Here and below we assume that the total number of decaying $B\bar B$ pairs equals $1/\Gamma^2$. Note that the integrand in (\ref{eq:nppoddfull}) vanishes at $\Delta t = 0$ ($t_1 = t_2$). This happens since at the moment of decay of the first $B$-meson, the flavor of the second one is opposite, therefore the simultaneous production of dileptons of the same sign is forbidden.

The ratio of the number of decays into leptons of the same signs to the number of decays into leptons of the opposite signs in the case of a $C$-odd wave function is equal to
\begin{equation}\label{eq:rodd}
	R_{\text{odd}} = \frac{N_{++} + N_{--}}{N_{+-} + N_{-+}} = \frac{x^2}{2 + x^2}.
\end{equation}

\subsection{$C$-even wave function}

For a wave function (\ref{func:even}), which is even with respect to the permutation of $B^0$ and $\overline{B}$$^0$, we obtain
\begin{eqnarray}
	N_{++} = N_{--} = \frac{1}{2} \int\limits_{0}^{+\infty} dt_1 \int\limits_{0}^{+\infty} dt_2 e^{- 2\Gamma t} \bigg[ \cos\bigg(\frac{\Delta m}{2}t_1\bigg) \sin\bigg(\frac{\Delta m}{2}t_2\bigg) + \nonumber\\ + \sin\bigg(\frac{\Delta m}{2}t_1 \bigg) \cos\bigg( \frac{\Delta m}{2}t_2\bigg) \bigg]^2,
\end{eqnarray}
\begin{eqnarray}
	N_{+-} = N_{-+} = \frac{1}{2} \int\limits_{0}^{+\infty} dt_1 \int\limits_{0}^{+\infty} dt_2 e^{-2\Gamma t} \bigg[ \cos\bigg(\frac{\Delta m}{2}t_1 \bigg) \cos\bigg(\frac{\Delta m}{2}t_2 \bigg) - \nonumber\\ -
	\sin\bigg(\frac{\Delta m}{2}t_2 \bigg) \sin\bigg(\frac{\Delta m}{2}t_1 \bigg) \bigg]^2.
\end{eqnarray}

Changing integration variables according to (\ref{eq:measure}) and integrating over $t$, we obtain:
\begin{eqnarray}\label{eq:nppevenfull}
	N_{++} = N_{--} = \frac{1}{4\Gamma(1 + x^2)}\int\limits_{0}^{+\infty} d(\Delta t) e^{- \Gamma \Delta t} \bigg( 1 + x^2 -\cos\Delta m \Delta t + \nonumber\\ + x \sin\Delta m \Delta t \bigg) =  \frac{1}{4 \Gamma^2} \frac{3x^2 + x^4}{(1 + x^2)^2}, 
\end{eqnarray}	
\begin{eqnarray}\label{eq:npmevenfull}
	N_{+-} = N_{-+} = \frac{1}{4\Gamma(1 + x^2)} \int\limits_{0}^{+\infty} d(\Delta t) e^{- \Gamma \Delta t} \bigg( 1 + x^2 +\cos\Delta m \Delta t - \nonumber\\ - x \sin\Delta m \Delta t \bigg) = \frac{1}{4 \Gamma^2} \frac{2 + x^2 + x^4}{(1 + x^2)^2}.
\end{eqnarray}
The integrand for $N_{++}$ and $N_{--}$ vanishes for $t_1 = t_2 = 0$, since at the moment of production the system consists of a pair $B^0 \overline{B}$$^0$, while at all subsequent moments the final state may contain leptons of the same and opposite signs.

The ratio of the number of decays into leptons of the same signs to the number of decays into leptons of the opposite signs in the case of a $C$-even wave function is equal to
\begin{equation}\label{eq:reven}
	R_{\text{even}} = \frac{N_{++} + N_{--}}{N_{+-} + N_{-+}} = \frac{3x^2 + x^4}{2 + x^2 + x^4}.
\end{equation}

\subsection{Incoherent production of $B$-mesons}

Performing analogous calculations for the wave function (\ref{func:inc}) we obtain
\begin{eqnarray}
	N_{++} = \int\limits_{0}^{+\infty} dt_1 \int\limits_{0}^{+\infty} dt_2 e^{-2\Gamma t} \bigg[ \cos\bigg(\frac{\Delta m}{2} t_1\bigg)  \sin\bigg(\frac{\Delta m}{2} t_2\bigg)\bigg]^2,\\
	N_{--} = \int\limits_{0}^{+\infty} dt_1 \int\limits_{0}^{+\infty} dt_2 e^{-2\Gamma t} \bigg[ \cos\bigg(\frac{\Delta m}{2} t_2\bigg)  \sin\bigg(\frac{\Delta m}{2} t_1\bigg)\bigg]^2,\\
	N_{+-} = \int\limits_{0}^{+\infty} dt_1 \int\limits_{0}^{+\infty} dt_2 e^{-2\Gamma t} \bigg[ \cos\bigg(\frac{\Delta m}{2} t_1\bigg) \cos\bigg(\frac{\Delta m}{2} t_2\bigg) \bigg]^2,\\
	N_{-+} = \int\limits_{0}^{+\infty} dt_1 \int\limits_{0}^{+\infty} dt_2 e^{-2\Gamma t} \bigg[ \sin\bigg(\frac{\Delta m}{2} t_1\bigg) \sin\bigg(\frac{\Delta m}{2} t_2\bigg) \bigg]^2.
\end{eqnarray}

Changing the integration variables according to (\ref{eq:measure}) and integrating over $t$, we get
\begin{eqnarray}
	N_{++} = N_{--} = \int\limits_{0}^{+\infty} d(\Delta t) \frac{e^{-\Gamma \Delta t}}{8 \Gamma (1 + x^2)}\bigg( 2 + 2x^2 - (2 + x^2)\cos\Delta m\Delta t + \nonumber \\ + x \sin \Delta m \Delta t\bigg) = \frac{1}{4 \Gamma^2} \frac{2x^2 + x^4}{(1 + x^2)^2},\\
	N_{+-} = \int\limits_{0}^{+\infty} d(\Delta t) \frac{e^{-\Gamma \Delta t}}{8 \Gamma (1 + x^2)(4 + x^2)}\bigg(2(1 + x^2)(8 + x^2) + (16 + 14x^2 + \nonumber \\ + x^4)\cos \Delta m \Delta t - x(8 + 5x^2)\sin\Delta m \Delta t\bigg) = \frac{1}{4 \Gamma^2}\frac{(2 + x^2)^2}{(1 + x^2)^2},\\
	N_{-+} = \int\limits_{0}^{+\infty}d(\Delta t) \frac{x^2e^{- \Gamma \Delta t}}{8 \Gamma (1 + x^2)(4 + x^2)}\bigg( 2 + 2x^2 + (x^2 - 2)\cos\Delta m \Delta t + \nonumber \\ + 3 x \sin \Delta m \Delta t \bigg) = \frac{1}{4 \Gamma^2}\frac{x^4}{(1 + x^2)^2}.
\end{eqnarray}

Thus, for the case of incoherent production of $B$-mesons we have
\begin{equation}\label{eq:rinc}
	R_{\text{inc}} = \frac{N_{++} + N_{--}}{N_{+-} + N_{-+}} = \frac{2 x^2 + x^4}{2 + 2x^2 + x^4}.
\end{equation}

\section{The case of one charged $B$-meson}

Let us consider the decay $\mathit{\Upsilon}(5S) \rightarrow B^+ \overline{B}$$^0 \pi^-$ with subsequent semileptonic decays $B^+ \rightarrow \ell^+ \nu_{\ell} X$ and either $\overline{B}$$^0 \rightarrow \ell^- \overline{\nu_{\ell}} \overline{X}$, or $\overline{B}$$^0 \rightarrow B^0 \rightarrow \ell^+ \nu_{\ell} X$. Let $N_{+}$ denote the number of events with both leptons of the same signs, and $N_{-}$ denote the number of events with both leptons of opposite signs. In all further calculations we substitute $\Gamma_{B^0} = \Gamma_{B^+} \equiv \Gamma$ (see footnote after equation (1)). From (\ref{eq:Bosc}) we readily get:
\begin{eqnarray}
	N_{+} = \int\limits_{0}^{+\infty}dt_1 \int\limits_{0}^{+\infty}dt_2 e^{- \Gamma (t_1 + t_2)} \sin^2\bigg( \frac{\Delta m t_2}{2}\bigg), \\
	N_{-} = \int\limits_{0}^{+\infty}dt_1 \int\limits_{0}^{+\infty}dt_2 e^{- \Gamma (t_1 + t_2)} \cos^2\bigg( \frac{\Delta m t_2}{2}\bigg).
\end{eqnarray}
With the help of (\ref{eq:measure}) we obtain
\begin{eqnarray}\label{eqq:np}
	N_{+} = \int\limits_{0}^{+\infty} d\Delta t \int\limits_{\Delta t/2}^{+\infty}dt e^{-2 \Gamma t} \bigg\{\sin^2\bigg[\frac{\Delta m}{2}\bigg( t - \frac{\Delta t}{2} \bigg)\bigg] + \sin^2\bigg[\frac{\Delta m}{2}\bigg( t + \frac{\Delta t}{2} \bigg)\bigg] \bigg\} = \nonumber \\ = \frac{1}{2}\int\limits_{0}^{+\infty} d\Delta t \frac{e^{-\Gamma \Delta t}}{\Gamma(4 + x^2)}\bigg(2 + x^2 - 2 \cos\Delta m \Delta t + x \sin\Delta m\Delta t\bigg) = \frac{1}{2 \Gamma^2} \frac{x^2}{1 + x^2},\\ \label{eqq:nn}
	N_{-} = \int\limits_{0}^{+\infty} d\Delta t \int\limits_{\Delta t/2}^{+\infty}dt e^{-2 \Gamma t} \bigg\{\cos^2\bigg[\frac{\Delta m}{2}\bigg( t - \frac{\Delta t}{2} \bigg)\bigg] + \cos^2\bigg[\frac{\Delta m}{2}\bigg( t + \frac{\Delta t}{2} \bigg)\bigg] \bigg\}= \nonumber \\ = \frac{1}{2} \int\limits_{0}^{+\infty} d\Delta t \frac{e^{-\Gamma \Delta t}}{\Gamma(4 + x^2)}\bigg(6 + x^2 + 2 \cos\Delta m \Delta t - x \sin \Delta m \Delta t\bigg) = \frac{1}{2 \Gamma^2} \frac{2 + x^2}{1 + x^2}.
\end{eqnarray}


\end{document}